\documentclass[twocolumn,showpacs,preprintnumbers,amsmath,amssymb,
superscriptaddress,nofootinbib]{revtex4}
\usepackage{graphicx}

\newcommand{\eps}{\varepsilon}
\newcommand{\tens}[1]{{\boldsymbol{#1}}}

\newcommand{\be}{\begin{equation}}
\newcommand{\ee}{\end{equation}}
\newcommand{\ba}{\begin{eqnarray}}
\newcommand{\ea}{\end{eqnarray}}

\newcommand{\hh}{\, ,\hspace{0.5cm}}

\newcommand{\hook}{\raisebox{-0.35ex}{\makebox[0.6em][r]
{\scriptsize $-$}}\hspace{-0.15em}\raisebox{0.25ex}{\makebox[0.4em][l]{\tiny
 $|$}}}
\newcommand{\eq}[1]{(\ref{#1})}

\newcommand{\n}[1]{\label{#1}}
\newcommand{\bs}[1]{{\boldsymbol{#1}}}
\newcommand{\nbs}[1]{{{#1}}}
\newcommand{\nbsi}[2]{{}^{#1}\!{{#2}}}
\newcommand{\bsi}[2]{{}^{\tiny #1}\!{\boldsymbol{#2}}}

\newcommand{\bsn}[2]{{}^{\tiny #2}\!{\boldsymbol{#1}}}
\newcommand{\bsnn}[2]{{}^{\tiny #2}\!\bar{{\boldsymbol{#1}}}}

\newcommand{\fb}[3]{{}_{\tiny #1}\!{\boldsymbol{#2}}^{\hat{#3}}}
\newcommand{\vb}[3]{{}^{#1}\!{\boldsymbol{#2}}_{\hat{#3}}}
\newcommand{\vnb}[3]{{}^{#1}\!{{#2}}_{\hat{#3}}}

\newcommand{\CQG}[3]{ {Classical Quantum Gravity\ }{\bf #1,\ } #3 (#2)} 

\newcommand{\PRL}[3]{ {Phys. Rev. Lett.\ }{\bf #1,\ } #3\ (#2)}

\begin{document}

\title{Solving parallel transport equations in the higher-dimensional
Kerr-NUT-(A)dS spacetimes}

\author{Patrick Connell}

\email{pconnell@phys.ualberta.ca}

\affiliation{Theoretical Physics Institute, University of Alberta, Edmonton,
Alberta, Canada T6G 2G7}

\author{Valeri P. Frolov}

\email{frolov@phys.ualberta.ca}

\affiliation{Theoretical Physics Institute, University of Alberta, Edmonton,
Alberta, Canada T6G 2G7}

\author{David Kubiz\v n\'ak}

\email{kubiznak@phys.ualberta.ca}

\affiliation{Theoretical Physics Institute, University of Alberta, Edmonton,
Alberta, Canada T6G 2G7}

\date{March 22, 2008}  

\begin{abstract} We obtain and study the equations describing the parallel
transport of  orthonormal frames along geodesics in a spacetime
admitting a non-degenerate principal conformal Killing--Yano tensor
$\bs{h}$.  We demonstrate that  the operator $\bs{F}$, obtained by a
projection of $\bs{h}$ to a subspace orthogonal to the velocity, has
in a  generic case eigenspaces of dimension not
greater than 2. Each of these eigenspaces are independently
parallel-propagated.  This allows one to reduce the parallel
transport equations to a set of the first order ordinary differential
equations  for the angles of rotation in the 2D eigenspaces. General
analysis is illustrated by studying the equations of the parallel
transport in the Kerr-NUT-(A)dS metrics. Examples of three, four, and
five dimensional Kerr-NUT-(A)dS are considered and it is shown that
the obtained  first order equations  can be solved by a separation of
variables.
\end{abstract}

\pacs{04.70.Bw, 04.50.+h, 04.20.Jb \hfill  Alberta-Thy-04-08}


\maketitle

\section{Introduction} 

Higher-dimensional black hole solutions of Einstein equations attract
a lot of interest. The most general known solution for an isolated
rotating higher-dimensional black hole in an asymptotically (anti) de
Sitter background, the Kerr-NUT-(A)dS metric, was obtained recently 
by Chen, L\"u, and Pope \cite{CLP}. In many respects this spacetime
is similar to the 4D Kerr-NUT-(A)dS solution. In particular, it possesses
hidden symmetries generated by the Killing--Yano (KY) and Killing
tensors. Namely, it admits a principal conformal Killing-Yano (CKY)
tensor \cite{FK,KF} which generates a tower of the Killing--Yano and
Killing tensors \cite{PKVK, KKPF}. The existence of the complete set
of irreducible Killing tensors and corresponding integrals of motion
makes the particle geodesic equation completely integrable
\cite{PKVK, KKPV}. For the same reason the Hamilton--Jacobi,
Klein--Gordon, and Dirac equations allow a separation of variables 
\cite{FKK,oy}. For more details on these results see recent reviews
\cite{review1, review2} and references therein.

One of the additional remarkable properties of the 4D Kerr metric, 
discovered by Marck in 1983 \cite{Marck, M2}, is that  the equations of
parallel transport can be  integrated. 
This result allows a
generalization: A parallel-propagated frame along a geodesic can be
constructed explicitly in an arbitrary 4D spacetime admitting the
rank-2 Killing--Yano tensor \cite{KM}.

Solving the parallel transport equations is useful for many problems
whose study is the behavior of extended objects moving in the Kerr and more
general geometries. In particular, it facilitated the study of tidal
forces acting on a moving body, for example a star, in the background
of a massive black hole (see, e.g., \cite{lm, Laguna,
FKNP,DFKNP,Siba:96,Siba:05}). In the special case of null geodesics
(rays) a solution to the parallel transport equations is required for
studying the propagation of light polarization
\cite{StCo:77,CoSt:77,CoPiSt:80}. 
In quantum physics the parallel
transport of frames is an important technical element of the point
splitting method which is used for calculation of renormalized values
of local observables (such as vacuum expectation values of currents,
stress-energy tensor etc.) in a curved spacetime. Solving of the
parallel transport equations is useful especially when  fields with
spin are considered (see, e.g., \cite{CHRIS}).

The purpose of the present paper is to generalize the results
\cite{Marck,KM} to the case of spacetimes with arbitrary number of 
dimensions admitting a non-degenerate principal CKY tensor. It was
demonstrated in \cite{o1} that the existence of such a CKY tensor
obeying the additional condition implies complete integrability  of 
geodesic motion. This `special' CKY tensor also distinguishes uniquely 
the Kerr-NUT-(A)dS spacetimes among all the Einstein spaces
\cite{o2}. However, in the first part of the paper  we do not limit
our  analysis to these metrics  and consider a more general
situation of a spacetime  with a closed 2-form of the 
non-degenerate principal CKY tensor $\tens{h}$ 
and an integrable particle geodesic motion. 

Let us outline the main idea of our construction.
Any 2-form  determines what is called a {\em Darboux basis},
that is a basis in which it has a simple standard canonical form. We
call the Darboux basis of $\bs{h}$ a {\em principal} basis.
For a strictly non-degenerate $\tens{h}$ the Darboux subspaces
are two-dimensional\footnote{In an odd number of spacetime dimensions 
there exists an additional one-dimensional 
zero-eigenvalue Darboux subspace of $\tens{h}$.}. It means that the `local' 
Darboux basis, defined in the tangent space of any spacetime point, is determined
up to 2D rotations in the Darboux subspaces. The union of local Darboux bases
of $\bs{h}$ forms a global principal
basis in the tangent bundle of the spacetime  manifold. 
In the case of the
Kerr-NUT-(A)dS metrics there exists a global principal basis in which
the Ricci rotation coefficients are simplified. We call this basis 
a {\em canonical} principal basis. 

Consider now a timelike geodesic describing the motion of a particle 
with velocity $\bs{u}$. We focus our attention on  a special
2-form  $\bs{F}$---obtained by a projection of the principal CKY
tensor $\bs{h}$  to a subspace orthogonal to the velocity $\bs{u}$. 
$\tens{F}$ has its own Darboux basis, which we call {\em comoving}. 
For any chosen geodesic the comoving basis is determined along its
trajectory. It can be easily shown that $\tens{F}$  is
parallel-transported along the geodesic. In particular, it means that
its eigenvalues and its Darboux subspaces, which we call the {\em
eigenspaces} of  $\tens{F}$, are parallel-transported.   We shall
show that for {\em generic} geodesics the eigenspaces of $\bs{F}$ are
at most 2-dimensional. In fact, the eigenspaces with non-zero
eigenvalues are 2-dimensional,  and the zero-value eigenspace is
1-dimensional for odd number  of spacetime dimensions and
2-dimensional for even. So, the comoving basis is defined up to
rotations  in each of the 2D eigenspaces. The 
{\em parallel-propagated} basis is a special comoving basis.  It can be found by
solving a set of the first order ordinary differential equations for
the angles of rotation in the 2D eigenspaces. 

For special geodesic trajectories the 2-form  $\bs{F}$ may become
{\em degenerate}, that is at least one of its eigenspaces will have
more than 2 dimensions. We shall demonstrate that the eigenspaces
with non-vanishing eigenvalues in such a degenerate case may be
4-dimensional.  In the odd number of spacetime dimensions one may
also have a 3-dimensional eigenspace with a zero eigenvalue. Nevertheless,
in these degenerate cases one can also obtain the parallel-transported basis 
by (now rather more complicated) time dependent rotations of the
comoving basis.

The paper is organized as follows. In Section~II we introduce the
form $\bs{F}$ and construct its Darboux basis. In Section~III we use
this comoving basis to derive the equations  for the parallel
transport of frames.  In Section~IV we adopt general results to the
case of the Kerr-NUT-(A)dS metrics. Concrete examples of
parallel-transported frames in $D=3, 4, 5$  Kerr-NUT-(A)dS spacetimes
are described in Section~V.  Section~VI  contains discussion.

\section{Hidden symmetries and comoving basis}

\subsection{Operator $\bs{F}$}

Consider  a  $D$-dimensional spacetime $M^D$ with the metric
\be\n{1}
\tens{g}=g_{ab}\tens{d}x^{a}\tens{d}x^{b}\, .
\ee
We assume that this metric has the signature $(-,+,\ldots,+)$. We also
assume that the spacetime possesses a closed CKY 2-form $\bs{h}$:
\ba
{\tens h}&=&{1\over 2}\,h_{ab}\, 
\tens{d}x^a\!\wedge \tens{d}x^b\,,\\
\nabla_{X}\bs{h}&=&-{1\over D-1}\,\bs{X}^{\flat}\wedge
\tens{\delta h}\, .\n{ccky}
\ea 
Here $\bs{X}$ is an arbitrary vector, $\tens{X}^{\flat}$ is the
corresponding form which has components
$(X^{\flat})_a=g_{ab}X^b$.  An inverse to $\flat$ operation is
denoted by $\sharp$. Namely if $\tens{\alpha}$ is a 1-form  then
$\bs{\alpha}^{\sharp}$ denotes a  vector with components 
$({\alpha}^{\sharp})^a=g^{ab}\alpha_b$. $\tens{\delta}$ denotes the
co-derivative. For a $p$-form  $\bs{\alpha}_p$ one has
$\tens{\delta\alpha}_p= \epsilon
\tens{*}\tens{d}\tens{*}\tens{\alpha}_p\,,$ where
$\epsilon=(-1)^{p(D-p)+p-1}$, and $\tens{*}$ denotes the Hodge star
operator. In usual tensor notations the definition \eq{ccky} of the
closed CKY tensor $\tens{h}$ reads
\be
\nabla_{c} h_{ab}=2g_{c[a}\xi_{b]},\quad
\xi_b=\frac{1}{D-1}\nabla_dh^{d}_{\ b}\,.
\ee
We call $\bs{h}$ a {\em principal} conformal Killing-Yano tensor. Its
dual tensor $ \bs{*h}$ is the Killing--Yano tensor ($(D-2)$-form).

Our aim is to construct a parallel-propagated frame along geodesic
in a spacetime with such a principal CKY tensor. First of all, we
construct a  {\em comoving} basis determined by the
principal CKY tensor and a time-like unit vector (vector of
velocity)\footnote{The construction we describe can be easily adapted
for the parallel transport of frames along spacelike geodesics.}.
At this stage our construction is local. We consider a tangent space
$T$ at a chosen point $p_0$. We denote by  $\bs{u}$ a chosen
time-like unit velocity vector $(u^a u_a\!=\!-1$), by $U$ a 1-dimensional space
generated by $\bs{u}$, and by  $V$  the $(D-1)$-dimensional subspace 
orthogonal to ${\bs{u}}$.  Thus $T$ is a direct sum of two orthogonal
subspaces, $U$ and $V$,
\be
T=U\oplus V\, .
\ee
Let  $\bs{F}$ be a 2-form defined by the relation (cf.~\cite{PKVK, KKPF, KKPV})
\ba\n{fh}
\bs{F}&=&\bs{h}+\bs{u}^{\flat}\wedge \bs{s}\, ,\n{Fop}\\
\bs{s}&=&\bs{u}\hook \bs{h}\hh s_b=u^a h_{ab}\, ,\n{ss}
\ea
or in components
\be
F_{ab}=h_{ab}+u_{a}u^{c} h_{cb}+h_{ac}u^{c}u_{b}
  =P_a^c P_b^d h_{cd}\, .
\ee
Here ${P_a^b=\delta_a^b+u^bu_a}$ is the projector to $V$.
$P_{ab}=g_{ab}+u_a u_b$ can be also  understood as a {\em positive
definite} metric in $V$ induced by its embedding into $T$. $F_{ab}$ and
$F^a_{\ b}=g^{ac}F_{cb}$ can be considered  as a 2-form and an
operator, respectively,  either in the subspace $V$ or in the
complete tangent space $T$. Since $F^a_{\ b}u^b=0$, the vector $\bs{u}\in T$
is an eigenvector of $\bs{F}$ with a zero eigenvalue.

\subsection{Comoving basis}

We demonstrate now that there exists such an orthonormal
basis in $V$ in which the operator $\bs{F}$ has the (matrix) form (see e.g.
\cite{Pras})
\be\n{F}
\mbox{diag}(0,\ldots,0,{\Lambda}_1,\ldots,{\Lambda}_p)\, ,
\ee
where ${\Lambda}_{\mu}$ are matrices of the form
\be
\Lambda_{\mu}=\left(   
\begin{array}{cc}
0 & \lambda_{\mu} {I}_{\mu}\\
-\lambda_{\mu} {I}_{\mu}& 0
\end{array}
\right)\, ,
\ee
and ${I}_{\mu}$ are unit matrices.  Such a basis, known as the {\em
Darboux basis}, can be constructed for any 2-form by a modified
version of the Gram--Schmidt process. We recall this procedure,
mainly, in order to fix the notations we shall use later in the
paper.

To construct the Darboux basis let us define an operator $\bs{F}^2$
with the components
\be\n{SS}
(F^2)^a_b=F^a_{\ \, c}F^c_{\ \, b}=P^a_d h^d_{\ e}P^e_f h^f_{\ k}P^k_b\, .
\ee
It annihilates the vector $\bs{u}$ and hence it can be considered as
an operator in $V$. Consider the following eigenvalue
problem
\be\n{Sv}
\tens{F}^2\tens{v}=-\lambda^2 \tens{v}\, .
\ee
Then we find 
\be\label{Fv}
\lambda^2 \bs{v}^2=-(\bs{v},\tens{F}^2\bs{v})=(\bs{F}\bs{v})^2\ge 0\, .
\ee
Here and later we denote by $\tens{F}^2\tens{v}$ a vector 
with components $(F^2)^{a}_b v^b$ and similarly
$\bs{F}\bs{v}$ a vector $F^{a}_{\ \,b}v^b$. 
Equation \eq{Fv} implies that $\lambda^2$ is non-negative. For
non-vanishing eigenvalues it is convenient to choose all $\lambda$ to be
positive. We enumerate $\lambda$ by index ${\mu}=0,\ldots,p$ and order
them as follows
\be\n{ll}
0=\lambda_0<\lambda_1<\ldots<\lambda_p\, .
\ee
If $\tens{F}^2$ does not have a zero eigenvalue, the first term
$\lambda_0$  in \eq{ll} is omitted. 
We denote  by $V_{{\mu}}$ a
subspace spanned by the eigenvectors of $\tens{F}^2$ corresponding to
$\lambda_{\mu}$.
These (Darboux) subspaces $V_\mu$ possess the property
\begin{equation}
\tens{F}V_\mu=V_\mu\,.
\end{equation}
We call them the {\em eigenspaces} of $\tens{F}$. 
Similarly we call $\lambda_\mu$ the `{\em eigenvalues}' of 
$\tens{F}$.\footnote{We put `\ ' marks to indicate that, strictly speaking,
$\lambda_\mu$ are eigenvalues not of $\tens{F}$ but of $\tens{F}^2$.}
The eigenspaces with different eigenvalues are mutually orthogonal  and
their direct sum forms the space $V$: 
\be
V=V_0\oplus V_{1} \oplus \ldots \oplus V_{p}\, . 
\ee

If $\lambda=0$ and the
corresponding subspace $V_0$ has $q_0$ dimensions, we denote an 
orthonormal basis in $V_0$ by
\be\label{V0}
\{\bsi{1}{n}_{\hat 0},\ldots,\bsi{q_0}{n}_{\hat 0}\}\, .
\ee

Let $\lambda \ne 0$. Consider a unit vector $\bsn{n}{1}$ from
$V_{\lambda}$ and denote $\bsnn{n}{1}=-\lambda^{-1}\tens{F}\,\bsn{n}{1}$.
One has 
\ba
(\bsn{n}{1},\bsnn{n}{1})&=&-\lambda^{-1}(\bsn{n}{1},\tens{F}\,\bsn{n}{1})=0\,
,\nonumber\\
\tens{F}\,\bsnn{n}{1}&=&-\lambda^{-1}\tens{F}^2\,\bsn{n}{1}=\lambda \bsn{n}{1}\, ,\\
(\bsnn{n}{1},\bsnn{n}{1})&=&\lambda^{-1}(\bsn{n}{1},\tens{F}\,\bsn{\bar{n}}{1})
=(\bsn{n}{1},\bsn{n}{1})=1\, .\nonumber
\ea
If $V_{\lambda}$ has 2 dimensions, $\{\bsn{n}{1},\bsnn{n}{1}\}$ is an
orthonormal basis in it. If $V_{\lambda}$ has more than 2 dimensions
we choose another unit vector $\bsn{n}{2}\in V_\lambda$ orthogonal to
both $\bsn{n}{1}$ and $\bsnn{n}{1}$ and denote
$\bsnn{n}{2}=-\lambda^{-1}\tens{F}\,\bsn{n}{2}$. Evidently, $\bsnn{n}{2}$
is orthogonal to $\bsn{n}{2}$ and has a unit norm. One also has
\ba
(\bsnn{n}{2},\bsn{n}{1})\!&=&\!\lambda^{-1}(\bsn{n}{2},\tens{F}\,\bsn{n}{1})=-
(\bsn{n}{2},\bsnn{n}{1})= 0\,, \nonumber\\
(\bsnn{n}{2},\bsn{\bar n}{1})\!&=&\!\lambda^{-1}(\bsn{n}{2},\tens{F}\,\bsnn{n}{1})=
(\bsn{n}{2},\bsn{n}{1})= 0\,.
\ea
Thus the vectors $\{\bsn{n}{2},\bsnn{n}{2}\}$ can be added to the set
$\{ \bsn{n}{1},\bsnn{n}{1}\}$.   If the dimension of $V_{\lambda}$ is
4 we already got an orthonormal basis in it. If the dimension is
greater than 4 the procedure must be repeated until finally a complete basis
is constructed. For the eigenvalue $\lambda=\lambda_{\mu}$ 
such a basis in $V_{{\mu}}$  is
\be\n{bas}
\{ \bsi{{1}}{n}_{\hat{\mu}},\bsi{{1}}{\bar{n}}_{\hat{\mu}}\ldots, 
\bsi{q_{\mu}}{n}_{\hat{\mu}},
\bsi{{q}_{\mu}}{\bar{n}}_{\hat{\mu}}\}\, . 
\ee
It is evident that each $V_{{\mu}}$ is an even dimensional
space; we denote its dimension by $2q_{\mu}$.  
We further denote by
\be\n{bas*}
\{\fb{1}{\varsigma}{{0}},\ldots,
\fb{{q_{0}}}{\varsigma}{0}\}\,,\ 
\{\fb{1}{\varsigma}{{\mu}},\fb{{1}}{\bar{\varsigma}}{{\mu}}
\ldots, \fb{{q_{\mu}}}{\varsigma}{\mu},
\fb{{q}_{\mu}}{\bar{\varsigma}}{{\mu}}\}\,,
\ee
bases of forms dual to the constructed orthonormal vector bases \eqref{V0}, 
\eqref{bas}.  These forms give bases in the cotangent spaces
$V_0^*$ and $V_\mu^*$.
We combine the bases \eqref{V0}, \eq{bas} (\eq{bas*}) with ${\mu}=0,\ldots,p$ to obtain
a complete orthonormal basis of vectors (forms) in the space
$V\,(V^*)$. The duality conditions read
\ba\n{bf}
&&\fb{s}{\varsigma}{{\mu}}(\bsi{s'}{n}_{\hat{\mu}'})=
\fb{s}{\bar{\varsigma}}{{\mu}}(\bsi{s'}{\bar{n}}_{\hat{\mu}'})=
\delta^{\mu}_{\mu'}\delta_{s}^{s'}\, ,\nonumber\\
&&\fb{s}{\varsigma}{{\mu}}(\bsi{s'}{\bar{n}}_{\hat{\mu}'})=
\fb{s}{\bar{\varsigma}}{{\mu}}(\bsi{s'}{{n}}_{\hat{\mu}'})=0\, .
\ea
Here for a given $\mu=0,\dots,p$ index $s$ takes the values $s=1,\dots,q_\mu$.
It is evident from the orthonormality of the constructed basis
that we also have
\begin{equation}\label{bstar}
\begin{split}
(\bsi{s}{n}_{\hat{\mu}})^\flat=\fb{s}{\varsigma}{{\mu}}\,,\quad
(\bsi{s}{\bar n}_{\hat{\mu}})^\flat=\fb{s}{\bar \varsigma}{{\mu}}\,,\\
(\fb{s}{\varsigma}{{\mu}})^\sharp=\bsi{s}{n}_{\hat{\mu}}\,,\quad
(\fb{s}{\bar \varsigma}{{\mu}})^\sharp=\bsi{s}{\bar n}_{\hat{\mu}}\,.
\end{split}
\end{equation}
In this basis  the antisymmetric operator $\bs{F}$ \eqref{Fop} takes the
form \eq{F}. If we consider $\bs{F}$ as an operator in
the complete tangent space $T$, the corresponding Darboux basis  is enlarged
by adding the vector $\bs{u}$ to it. In this enlarged basis the operator
$\bs{F}$ has the same form \eq{F}, with the only difference that now
the total number of zeros is not $q_0$, but $q_0+1$. 
To remind that
the constructed Darboux basis depends on the velocity $\bs{u}$ of a
particle and $\bs{u}$ is one of its elements we call this basis  {\em
comoving}. The characteristic property of the comoving frame is that
all spatial components of the velocity vanish. 

Although our construction was local, we can naturally extend
the comoving basis along the whole geodesic trajectory. 
In a general case, however, the constructed comoving frame is not parallel-propagated.
The parallel-propagated frame can be obtained by performing additional rotations in each
of the  parallel-propagated eigenspaces of $\bs{F}$. The equations
for the corresponding rotation angles will be derived in the next
section. Before we do that we demonstrate that 
for a strictly non-degenerate principal CKY tensor $\tens{h}$ the structure of the eigenspaces of $\tens{F}$,
and hence the comoving basis, significantly simplifies.

\subsection{Eigenspaces of $\tens{F}$}
To  treat both cases of the 
even and odd  dimensional spacetime $M^D$ simultaneously we denote 
\be
D=2n+\varepsilon\,,
\ee
where $\varepsilon=0$  and
$\varepsilon=1$ for  even and odd number of dimensions, respectively. 

In the comoving frame constructed above the form $\tens{F}$
reads:
\be\label{FF}
\bs{F}=\sum_{{\mu}=1}^{p} \lambda_{\mu} (\sum_{j=1}^{q_{\mu}} 
\fb{j}{\varsigma}{{\mu}}\wedge\fb{{j}}{\bar{\varsigma}}{{\mu}})\,.
\ee
It is convenient to call the ordered eigenvalues $\lambda_\mu$ \eqref{ll} together 
with their degeneracies a `{\em spectrum}' of $\tens{F}$ (cf. Eq. \eqref{Sv}).
We denote it by 
\be\label{spectrum}
S(\tens{F})=\{0, \underbrace{0,\dots,0}_{q_0},
\underbrace{\lambda_1,\dots,\lambda_1}_{2q_1},\dots,
\underbrace{\lambda_p,\dots,\lambda_p}_{2q_p}\}\,.
\ee 
The first zero eigenvalue corresponds to the 1-dimensional subspace $U$ spanned by
$\tens{u}$. One also has 
\begin{equation}\label{r}
D=1+q_0+2k\,,\quad k = \sum_{\mu=1}^pq_\mu\,.
\end{equation}

\subsubsection{Structure of $V_0$}

Let us now impose the condition that $\tens{h}$ is {\em non-degenerate}, that is its (matrix) rank is $2n$. Then one has 
\be\label{q0}
q_0=\left\{ \begin{array}{cc}
1\, ,  &\mbox{   for  }\varepsilon=0\, ,\\
0 \mbox{  or } 2\, , & \mbox{   for  }\varepsilon=1\, .
\end{array}
\right.
\ee
Let us prove this assertion.
For an arbitrary form $\tens{\alpha}$ we denote  
\be
\bs{\alpha}^{\wedge m}=\underbrace{\bs{\alpha}\wedge \ldots \wedge
\bs{\alpha}}_{\mbox{\tiny{total of m factors}}}\, .
\ee
From the definition  \eqref{Fop} of $\tens{F}$ we find
\be\n{hhh}
\bs{h}^{\wedge m}=\bs{F}^{\wedge m}-m\bs{F}^{\wedge (m-1)}\wedge
\bs{u}^{\flat}\wedge\bs{s}\, ,
\ee
where we have used the property of the external product  $ \bs{\alpha}_p\wedge
\bs{\alpha}_q=(-1)^{pq} \bs{\alpha}_q\wedge \bs{\alpha}_p\,$.
It is obvious from \eqref{FF} that the (matrix) rank of $\tens{F}$ is $2k$, that 
is $\tens{F}^{\wedge (k+1)}=0$. So, using \eqref{hhh} we have $\tens{h}^{\wedge (k+2)}=0$. It means that  
for a non-degenerate (matrix rank $2n$) $\tens{h}$ we have $k+2\ge n+1$.
Employing \eqref{r} this is equivalent to $q_0\le 1+\varepsilon$. This,  
together with the fact that $q_0$ has to be even for $D$ odd and vice versa,
proves \eqref{q0}. 

Let us now consider a nontrivial $V_0$, that is $V_0$ with
$q_0=1+\varepsilon$, $n-1=k$.
The vectors spanning it can be found as the eigenvectors of 
the operator $\tens{F}^2$ with zero eigenvalue, not belonging to $U$.
There is, however, a more direct way which was already used by Marck in 4D. Let us consider a Killing--Yano 
$(2+\varepsilon)$-form\footnote{It was demonstrated in \cite{KKPF} (see also \cite{review1, review2}) that an exterior product of closed CKY tensors is again a closed CKY tensor. The Hodge dual of a closed CKY tensor is a Killing--Yano tensor.} 
\be
\bs{f}=*\bs{h}^{\wedge k}\,,
\ee
and use it to define a $(1+\varepsilon)$-form
\be\label{z}
\bs{z}=\bs{u}\hook \bs{f}\, .
\ee
Using the relation
\be
\bs{X}\hook *\bs{\alpha}=*(\bs{\alpha}\wedge \bs{X}^{\flat})\, ,
\ee
and the equation \eqref{hhh} one obtains
\be\n{zzz}
\bs{z}=\bs{u}\hook *\bs{h}^{\wedge k}=*(\bs{h}^{\wedge
k}\wedge \bs{u}^{\flat})=
*(\bs{F}^{\wedge k}\wedge \bs{u}^{\flat})\, .
\ee
Employing \eqref{FF} we have
\be\n{FFF}
\bs{F}^{\wedge k}\!=B\,
\fb{1}{\varsigma}{1}\wedge\fb{1}{\bar{\varsigma}}{{1}}\!\wedge\! \ldots
\wedge\!\fb{{q}_p}{\bar{\varsigma}}{p}\,,
\ B=k!\!\prod_{\mu=1}^{p}\lambda_\mu^{q_\mu}\,.
\ee
This means that $\tens{z}$ spans $V_0^*$.
In the even number of spacetime dimensions the space 
$V_0^*$ is 1-dimensional and $\fb{\!}{\varsigma}{0}=\tens{z}/|z|$. 
Hence, using \eqref{bstar}, $\,\vb{\!}{n}{0}=\tens{z^\sharp}/|z|$ spans $V_0$. 
In the odd number of spacetime dimensions
\be
\bs{z}=\mbox{const} \ \fb{1}{\varsigma}{0}\wedge \fb{2}{\varsigma}{0}\, .
\ee
Hence, the 2-form $\bs{z}$ determines the orthonormal
basis $\{\vb{1}{n}{0} , \vb{2}{n}{0}\}$ in $V_0$ up to a 2D rotation.

Let us finally consider the odd dimensional case in more detail. Expanding the
characteristic equation for the operator $\bs{F}$ one has 
\be
0=\mbox{det}(\bs{F}-\lambda \tens{I})=
a(\bs{u})+b(\bs{u})\lambda^2+\ldots 
\ee
The condition that $q_0=2$ implies that
$a(\bs{u})=\mbox{det}(\bs{F})=0$. This imposes a constraint on $\bs{u}$.
It means that $q_0=2$ is a {\em degenerate} case which happens only 
for special trajectories $\tens{u}$. For a {\em generic} 
(not special) $\bs{u}$ one has trivial $V_0$ with $q_0=0$.

\subsubsection{Eigenspaces $V_\mu$}

We call $\bs{h}$ {\em strictly non-degenerate} if it is
non-degenerate and its eigenvalues are different, that is each of the
corresponding Darboux subspaces has not more than 2 dimensions. It is
possible to show (see Appendix A) that for a strictly non-degenerate
$\bs{h}$ the dimensionalities of the eigenspaces of $\bs{F}$ with
non-zero eigenvalues obey the inequalities $q_\mu\le 2$. The case of
$q_\mu= 2$ is possible only in a degenerate case when the vector
$\bs{u}$ obeys a special condition.

\section{Equations of parallel transport}
In this section we describe how to obtain the parallel-transported basis 
from the comoving basis constructed above.
Let $\gamma$ be a timelike geodesic and $u^a=dx^a/d\tau$ be a tangent
vector to it. 
We denote the covariant derivative of
a tensor $\tens{T}$ along $\gamma$ by 
\begin{equation}
\tens{\dot T} =\nabla_u \tens{T}=u^a \nabla_a \tens{T}\,.
\end{equation}
In particular $\tens{\dot u}=0$.
The crucial fact for our construction is that the form 
$\tens{F}$ is parallel-transported along the geodesic
$\gamma$ \cite{PKVK,KKPF}
\be
\tens{\dot F}=0\,. 
\ee
Let us prove this important property.
Using the definition \eqref{Fop} of $\tens{F}$, 
the fact that the velocity is normalized $\tens{u}\hook \tens{u}^\flat=-1$,
the definition \eqref{ss} of $\tens{s}$, 
the property of the hook operator
\begin{equation}
\tens{u}\hook (\bs{\alpha}_p\wedge\! \bs{\alpha}_q)=
(\tens{u}\hook \tens{\alpha}_p)\wedge\! \tens{\alpha}_q\!+\!
(-1)^{p} \bs{\alpha}_p (\tens{u}\hook \bs{\alpha}_q),
\end{equation}
and the defining equation \eqref{ccky} for the principal CKY tensor $\tens{h}$, we have 
\begin{equation}
\begin{split}
\tens{\dot F}=&\ \tens{\dot h}+\tens{u}^\flat\!\wedge \tens{\dot s}=
-(\tens{u}\hook\tens{u}^\flat)\,\tens{\dot h}+\tens{u}^\flat\!\wedge\!(\tens{u}\hook \tens{\dot h})\\
=&-\!\tens{u}\hook (\tens{u}^\flat\wedge\nabla_u\tens{h})\\
=&\,\frac{1}{D-1}\,\tens{u}\hook (\tens{u}^\flat\!\wedge\!\tens{u}^\flat\!\wedge\! \tens{\delta h})=0\,.
\end{split}
\end{equation}
Notice that since $\bs{F}$ is parallel-propagated along $\gamma$, any
object constructed from $\bs{F}$ and metric $\bs{g}$ is also
parallel-propagated. In particular, this is true for the operator $\tens{F}^2$ and its eigenvalues $\lambda_{\mu}$.
Hence, the eigenspaces $V_{\mu}$ 
are independently parallel-transported, that is
\begin{equation}\label{PTV}
\bs{\dot v}\in V_{\mu} \quad {\rm for} \ 
\forall\, \bs{v}\in V_{\mu}.
\end{equation}  
Indeed, using \eqref{Sv} we find
\begin{equation}
\tens{F}^2 \tens{\dot v}=\nabla_u\!\left(\tens{F}^2\tens{v}\right)=
\nabla_u\!\left(-\lambda_\mu^2\tens{v}\right)=
-\lambda_\mu^2 \tens{\dot v}\,,
\end{equation}
which proves \eqref{PTV}.
It should be emphasized that the dimension of an eigenspace of $\tens{F}$
is also constant along $\gamma$. 
For generic geodesics the eigenspaces of $\tens{F}$ with non-zero eigenvalues are always
2-dimensional, while the subspace with zero eigenvalue ($U\oplus V_0$) has $2-\varepsilon$ dimensions.
There might also exist a zero measure set of special geodesics for which either
an eigenspace of $\bs{F}$ with non-zero eigenvalue has not 2 but 4
dimensions or (in the odd dimensional case) the eigenspace of $\bs{F}$
with zero eigenvalue has 3 dimensions. (See Section II C.)

The parallel-propagated basis can be obtained from the comoving basis by time dependent rotations in the eigenspaces of $\tens{F}$. We
denote the corresponding matrix of rotations by $\nbs{O}(\tau)$.
Similar to $\bs{F}$ it has the following structure
\be
\nbs{O}=\mbox{diag}(\nbs{O}_{\hat{0}},\nbs{O}_{\hat{1}},\ldots,\nbs{O}_{\hat{p}})\,.
\ee

For $\lambda_{\mu}>0$,  $\nbs{O}_{\hat{\mu}}$ are $2q_{\mu}\times
2q_{\mu}$ orthogonal matrices.  Let $\{ \vb{s}{p}{\mu},
\vb{s}{\bar{p}}{\mu}\}$ be a parallel-propagated basis in the
eigenspace $V_{\mu}$ and  $\{ \vb{s}{n}{\mu}, \vb{s}{\bar{n}}{\mu}\}$
be the `original' comoving basis then
\be\n{tran}
\left( \begin{array}{c}
\vb{s}{p}{\mu}\\
\vb{s}{\bar{p}}{\mu}
\end{array}\right)
=\sum_{s'=1}^{q_\mu}\nbs{O}_{\hat{\mu}\ s'}^{\ \ s}
\left(
\begin{array}{c}
\vb{s'}{n}{\mu}\\
\vb{s'}{\bar{n}}{\mu}
\end{array}\right)\, .
\ee
Here, for fixed values $\{\hat{\mu}, s, s'\}$,  $\nbs{O}_{\hat{\mu}\ s'}^{\ \ s}$ are $2\times 2$ matrices.
Differentiating \eq{tran} along the geodesic and using the fact that
$\{ \vb{s}{p}{\mu}, \vb{s}{\bar{p}}{\mu}\}$ are parallel-propagated
one gets
\be
\sum_{s'=1}^{q_\mu}{\nbs{{\dot O}}}_{\hat{\mu}\ s'}^{\ \ s}
\left(\!
\begin{array}{c}
\vb{s'}{n}{\mu}\\
\vb{s'}{\bar{n}}{\mu}
\end{array}\!\right)=-\!\sum_{s'=1}^{q_\mu}{\nbs{{O}}}_{\hat{\mu}\ s'}^{\ \ s}
\left(\!
\begin{array}{c}
\vb{s'}{\dot{n}}{\mu}\\
\vb{s'}{\dot{\bar{n}}}{\mu}
\end{array}\!\right)\, .
\ee
This gives the following set of the first order differential equations for
$\nbs{O}_{\hat{\mu}\ s'}^{\ \ s}$ 
\be\n{dott}
{\nbs{{\dot O}}}_{\hat{\mu}\ s'}^{\ \ s}=-
\sum_{s''=1}^{q_\mu}{\nbs{{O}}}_{\hat{\mu}\ s''}^{\ \ s}
\nbs{N}_{\hat{\mu}\ s'}^{\ \ s''}
\, ,
\ee
where
\be\label{NN}
\nbs{N}_{\hat{\mu}\ s'}^{\ \ s''}=
\left(
\begin{array}{c c}
(\vb{s'}{\dot{n}}{\mu},\vb{s''}{n}{\mu}) &
(\vb{s'}{\dot{n}}{\mu},\vb{s''}{\bar{n}}{\mu})\\
(\vb{s'}{\dot{\bar{n}}}{\mu},\vb{s''}{n}{\mu}) &
(\vb{s'}{\dot{\bar{n}}}{\mu},\vb{s''}{\bar{n}}{\mu})
\end{array}
\right)\, .
\ee
For generic geodesics the parallel transport equations are
greatly simplified. In this case each of the eigenspaces $V_{\mu}$ 
is two-dimensional. 
The equations \eq{tran} take the form
\ba\label{pp}
\bs{p}_{\hat{\mu}}&=&\cos \beta_{\mu} \bs{n}_{\hat{\mu}}-
\sin \beta_{\mu} \bs{\bar{n}}_{\hat{\mu}}\, ,\nonumber\\
\bs{\bar{p}}_{\hat{\mu}}&=&\sin \beta_{\mu} \bs{n}_{\hat{\mu}}+
\cos \beta_{\mu} \bs{\bar{n}}_{\hat{\mu}}\, .
\ea
It is easy to check that the equations \eq{dott} reduce to the
following first order equations
\be\n{dot}
\dot{\beta}_{\mu}=(\bs{\dot {\bar{n}}}_{\hat{\mu}}, \bs{n}_{\hat{\mu}})
=-(\bs{n}_{\hat{\mu}},\bs{\dot {\bar{n}}}_{\hat{\mu}})\,.
\ee
If at the initial point $\tau=0$ bases $\{ \tens{p} \}$ and $\{ \tens{n}\}$ coincide,  
the initial conditions for the equations \eq{dot} are
\begin{equation}\label{ic}
\beta_{\mu}(\tau=0)=0\,.
\end{equation}

For $\lambda_0=0$, $\nbs{O}_{\hat{0}}$
is a $q_0\times q_0$ matrix. In even number of spacetime dimensions,
$q_0=1$, and $V_0$ is spanned by $\tens{n}_{\hat 0}$ which is already
parallel-propagated. Therefore we have 
$\nbs{O}_{\hat{0}}=1$. For odd number of spacetime dimensions,
$\nbs{O}_{\hat{0}}$ is present only in the degenerate case, 
$q_0=2$, that is when $V_0$ is spanned by $\{\vb{1}{n}{0} , \vb{2}{n}{0}\}$.
The parallel-propagated vectors $\{ \vb{1}{p}{0}, \vb{2}{p}{0}\}$
are then given by the analogue of the equations \eqref{pp}-\eqref{ic}.

\section{Parallel transport in Kerr-NUT-(A)dS spacetimes}

\subsection{Kerr-NUT-(A)dS spacetimes}

In a Kerr-NUT-(A)dS spacetime of arbitrary dimension ($D>2$) the
metric can be written in the form \cite{CLP}
\begin{equation}\label{metrics}
\tens{g}=\!\sum_{\mu=1}^{n-1} (\tens{\omega}^{\hat \mu} 
\tens{\omega}^{\hat \mu}\! + \tens{\bar \omega}^{\hat \mu} 
\tens{\bar \omega}^{\hat \mu}) + \tens{\omega}^{\hat n} 
\tens{\omega}^{\hat n} -\tens{\bar \omega}^{\hat n} 
\tens{\bar \omega}^{\hat n}\!
+ \eps\, \tens{\omega}^{\hat \epsilon} \tens{\omega}^{\hat \epsilon},
\end{equation}
where the basis 1-forms are 
\ba\label{one-forms}
\tens{\omega}^{\hat n} &=&\frac{\tens{d}r}{\sqrt{Q_n}}\,,\quad 
\tens{\omega}^{\hat \mu} = \frac{\tens{d}x_{\mu}}{\sqrt{Q_{\mu}}}\,,\quad 
\mu=1,\dots,n-1\,,\nonumber\\
\tens{\bar \omega}^{\hat \mu} &=& \sqrt{Q_{\mu}}
 \sum_{j=0}^{n-1}A_{\mu}^{(j)}\tens{d}\psi_j\;,\quad 
\mu=1,\dots,n\,,\nonumber\\
\tens{\omega}^{\hat \epsilon} &=&\sqrt{\frac{-c}{A^{(n)}}}
\sum_{j=0}^nA^{(j)}\tens{d}\psi_j\;.
\ea
Notice that we enumerate the basis $\{\tens{\omega}\}$ so that 
$\tens{\bar \omega}^{\hat n}$ is (the only one) timelike 1-form. 
Here, 
\ba\label{func}
A_{\mu}^{(j)}\!&=&\!\!\!\!\sum_{\ \substack{\nu_1<\dots<\nu_j\ \\
\nu_i\ne\mu}}\!\!\!\!\!x^2_{\nu_1}\dots x^2_{\nu_j},\quad\!\!\! 
A^{(j)}=\!\!\!\!\!\!\sum_{\nu_1<\dots<\nu_j}
\!\!\!\!\!x^2_{\nu_1}\dots x^2_{\nu_j}\;,\nonumber\\
Q_{\mu}\!&=&\!\,\frac{X_{\mu}}{U_{\mu}}\,,\ 
U_{\mu}=\prod_{\substack{\nu=1\\\nu\ne\mu}}^{n}(x_{\nu}^2-x_{\mu}^2)\,,\ 
x_n^2=-r^2\,,
\nonumber\\
X_{n}\!&=&\!-\sum\limits_{k=\varepsilon}^{n}c_k(-r^2)^k-2m r^{1-\varepsilon}
+\frac{\varepsilon c}{r^2}\,,\nonumber\\
X_{\mu}\!&=&\!\sum\limits_{k=\varepsilon}^{n}
c_kx_{\mu}^{2k}-2b_{\mu}x_{\mu}^{1-\varepsilon}
+\frac{\varepsilon c}{x_{\mu}^2}\,.
\ea
Time is denoted by $\psi_0$, azimuthal coordinates by $\psi_j$,
${j=1,\dots,D-n-1}$, $r$ is the Boyer-Lindquist type radial
coordinate, and ${x_\mu}$, ${\mu=1,\dots, n-1}$, stand for latitude
coordinates.  The parameter $c_n$ is proportional to the cosmological
constant \cite{HHOY} \begin{equation} {R_{ab}=(-1)^{n}(D-1)c_n\,
g_{ab}}\,, \end{equation} and the remaining constants $c_k$, $c>0$, and
$b_{\mu}$ are related to rotation parameters, mass, and NUT
parameters.

More generally, it is possible to consider a broader class of metrics 
\eqref{metrics} where $X_n(r), X_\mu(x_\mu)$, are arbitrary functions.
To stress that such metrics do not necessarily satisfy the Einstein equations we 
call them {\em off-shell} metrics.
All the statements and formulas  formulated for  
the Kerr-NUT-(A)dS solutions below
are also valid off-shell.

The principal CKY tensor reads \cite{KF}
\begin{equation}\label{KYKNA}
\tens{h}=\sum_{\mu=1}^{n-1} x_\mu \tens{\omega}^{\hat \mu}
\wedge \tens{\bar \omega}^{\hat \mu}
-r \tens{\omega}^{\hat n}\wedge \tens{\bar \omega}^{\hat n}\,.
\end{equation}
This means that basis $\{\tens{\omega}\}$ is a principal one. This
principal basis has an additional nice property; namely that many of the
Ricci coefficients of rotation vanish \cite{HHOY}. We call this
special principal basis {\em canonical}.  
The second-rank irreducible Killing tensors are \cite{KKPF, FKK}
($j=1,\dots, D-n-1$) \begin{equation}
\begin{split}
\tens{K}^{(j)}=&\,\sum_{\mu=1}^{n-1} A_{\mu}^{(j)} 
(\tens{\omega}^{\hat \mu} \tens{\omega}^{\hat \mu}+
\tens{\bar \omega}^{\hat \mu} \tens{\bar \omega}^{\hat \mu})\\
+&\, A_{n}^{(j)} (\tens{\omega}^{\hat n} \tens{\omega}^{\hat n} 
-\tens{\bar \omega}^{\hat n} \tens{\bar \omega}^{\hat n})\!
+ \eps  A^{(j)}  \tens{\omega}^{\hat \epsilon} \tens{\omega}^{\hat \epsilon} \, .
\end{split}
\end{equation}

The geodesic motion of a particle in the Kerr-NUT-(A)dS spacetime 
is completely integrable 
\cite{PKVK, KKPV} and its velocity reads \cite{KKPF, review2}:
\begin{equation}
\tens{u}^\flat=\sum_{\mu=1}^n \bigl( u_{\hat \mu} \tens{\omega}^{\hat \mu} + {\bar u}_{\hat \mu} \tens{\bar \omega}^{\hat \mu}\bigr)
+\eps\,u_{\hat \epsilon} \tens{\omega}^{\hat \epsilon}\,,
\end{equation}
where the vielbein components of the velocity are 
\begin{equation}\label{basisvectors}
\begin{aligned}
u_{\hat n} =&\, \frac{\sigma_n}{(X_n U_n)^{1/2}}\,\bigl(
W_n^2 -X_n V_n\bigr)^{1/2}\,,\\
u_{\hat \mu} =&\, \frac{\sigma_\mu}{(X_\mu U_\mu)^{1/2}}\,\bigl(
X_\mu V_\mu-W_\mu^2 \bigr)^{1/2}\,,\\
{\bar u}_{\hat n}=&\,\frac{W_n}{(X_n U_n)^{1/2}}\,,\ 
{\bar u}_{\hat\mu}=\frac{W_\mu}{(X_\mu U_\mu)^{1/2}}\,,\\
u_{\hat \epsilon} =&\,\frac{\Psi_n}{\sqrt{-cA^{(n)}}}\,. 
\end{aligned}
\end{equation} 
Here constants $\sigma_\mu=\pm 1$ ($\mu=1,\dots, n$) are independent 
one of another and we have defined 
\begin{equation}\label{VW}
\begin{split}
V_n=&-\!\!\sum_{j=0}^{m}r^{2(n-1-j)}\kappa_j\,,\ 
V_\mu\!=\!\sum_{j=0}^{m}(-x_{\mu}^2)^{n-1-j}\kappa_j\,,\\  
W_n=&-\!\!\sum_{j=0}^{m}r^{2(n-1-j)}\Psi_{\!j}\,,\ 
W_\mu\!=\!\sum_{j=0}^{m}(-x_{\mu}^2)^{n-1-j}\Psi_{\!j}\,.
\end{split}
\end{equation}
The quantities $\Psi_j$ and $\kappa_j$ are conserved and connected 
with the Killing vectors and the Killing tensors, respectively. The constant 
$\kappa_0$ denotes the normalization of the velocity 
$\kappa_0=u^a u_a=-1$ and 
\begin{equation}
\kappa_n=-\frac{\Psi_n^2}{c}\,.
\end{equation}

We shall construct a parallel-propagated frame for geodesic motion in 3
steps. 
At first we use the freedom of local rotations in the 2D Darboux spaces of $\tens{h}$
to introduce the {\em velocity adapted} principal basis in which 
$n$ components of the velocity vanish.
As the second step, by studying the eigenvalue problem for the
operator $\tens{F}^2$ we find a transformation connecting the velocity
adapted basis to a comoving basis. And finally, we derive the
equations for the rotation angles in the eigenspaces of $\bs{F}$
which transform the obtained comoving basis into the 
parallel-propagated one.

\subsection{Velocity  adapted principal basis}

To construct the velocity adapted principal basis we perform  the
boost transformation in the $\{\tens{\bar \omega}^{\hat n},
\tens{\omega}^{\hat n}\}$ 2-plane and the rotation transformations in
each of the $\{\tens{\bar \omega}^{\hat \mu}, \tens{\omega}^{\hat
\mu}\}$,  $\mu<n$, 2-planes:
\begin{equation}\label{newframe}
\begin{split}
\tens{\bar o}^{\hat n}=&\,\cosh\alpha_n 
\tens{\bar \omega}^{\hat n}+\sinh\alpha_n\tens{\omega}^{\hat n}\,,\\
\tens{o}^{\hat n}=&\,\sinh\alpha_n \tens{\bar \omega}^{\hat n}
+\cosh\alpha_n\tens{\omega}^{\hat n}\,,\\
\tens{\bar o}^{\hat \mu}=&\,\cos\alpha_\mu
\tens{\bar \omega}^{\hat \mu}+\sin\alpha_\mu \tens{\omega}^{\hat \mu}\,,\\
\tens{o}^{\hat \mu}=&\,-\sin\alpha_\mu
\tens{\bar \omega}^{\hat \mu}+\cos\alpha_\mu \tens{\omega}^{\hat \mu}\,,\\
\tens{o}^{\hat \epsilon}=&\ \tens{\omega}^{\hat \epsilon}\,.
\end{split}
\end{equation}
For arbitrary angles $\alpha_\mu$ ($\mu=1,\dots, n$) this transformation preserves the
form of the metric and of the principal CKY tensor:
\ba
\tens{g}\!&=&\!\sum_{\mu=1}^{n-1}(\tens{o}^{\hat \mu} \tens{o}^{\hat \mu}\! 
+\!\tens{\bar o}^{\hat \mu} \tens{\bar o}^{\hat \mu})\!
+\! \tens{o}^{\hat n} \tens{o}^{\hat n} \! -
\!\tens{\bar o}^{\hat n} \tens{\bar o}^{\hat n}\!+ 
\eps\, \tens{o}^{\hat \epsilon} \tens{o}^{\hat \epsilon},\nonumber\\
\tens{h}\!&=&\!\sum_{\mu=1}^{n-1} x_\mu 
\tens{o}^{\hat \mu}\wedge \tens{\bar o}^{\hat \mu}
-r \tens{o}^{\hat n}\wedge \tens{\bar o}^{\hat n}
\,.
\ea
Let us define
\begin{equation}
\begin{split}
{\bar v}_{\hat n}=&\,-\sqrt{{\bar u}_{\hat n}^2-u_{\hat n}^2}
=-\sqrt{\frac{V_n}{U_n}}\,,\\
{\bar v}_{\hat \mu}=&\,\sqrt{{\bar u}_{\hat\mu}^2+u_{\hat \mu}^2}
=\sqrt{\frac{V_\mu}{U_\mu}}\,.
\end{split}
\end{equation}
Then, specifying the values of $\alpha_\mu$ to be 
\begin{equation}\label{br}
\begin{split}
\cosh\alpha_n=&\,\frac{{\bar u}_{\hat n}}{{\bar v}_{\hat n}}\,,\quad
\sinh\alpha_n=\frac{u_{\hat n}}{{\bar v}_{\hat n}}\\
\cos\alpha_\mu=&\,\frac{{\bar u}_{\hat \mu}}{{\bar v}_{\hat \mu}}\,,\quad
\sin\alpha_\mu=\frac{u_{\hat \mu}}{{\bar v}_{\hat \mu}}\,,
\end{split}
\end{equation}
one obtains the following form of the velocity
\begin{equation}
\tens{u}^\flat=\sum_{\mu=1}^n {\bar v}_{\hat \mu}\tens{\bar o}^{\hat \mu}+
\eps u_{\hat \epsilon}\tens{o}^{\hat \epsilon}\,.
\end{equation} 
It means that after this transformation the velocity vector $\tens{u}$ has only 
$(n+\varepsilon)$ non-vanishing components. This simplifies considerably 
the construction of the comoving and the parallel-propagated bases.
Notice also that the boost in the $\{\tens{\bar \omega}^{\hat n},
\tens{\omega}^{\hat n}\}$ 2-plane is function  of $r$ only and the
rotation in each $\{\tens{\bar \omega}^{\hat \mu}, \tens{\omega}^{\hat
\mu}\}$ 2-plane is  function of $x_\mu$ only.  The components of
the velocity in the adapted basis $\{\tens{o}\}$ 
depend on constants $\kappa_j$ only; 
constants $\Psi_j$ and $\sigma_\mu$ are
absorbed in the definition of the new frame.

\subsection{Parallel-propagated frame}

It is obvious from the expression \eqref{KYKNA} that at a generic 
point of the manifold \eqref{metrics} the principal CKY tensor
$\tens{h}$ is strictly non-degenerate and we may use the theory
described in Sections II and III.
In particular, for generic geodesics  the operator 
$\tens{F}^2$ possesses twice degenerate non-zero eigenvalues, and the 
nontrivial eigenspace $V_0$, which is present only in even
dimensions, is 1-dimensional space determined by the 
properly normalized $\tens{z}^\sharp$.   Therefore 
the problem 
of finding the parallel-propagated frame in Kerr-NUT-(A)dS spacetimes 
reduces to  finding the eigenvectors $\{\tens{n}_{\hat \mu},
\tens{\bar n}_{\hat \mu}\}$ spanning  the 2-plane eigenspaces
$V_{\mu}$ and subsequent 2D rotations \eqref{pp} in these spaces. 

A degenerate case which requires a special consideration arises when initially different
elements of the spectrum $S(\tens{F})$, \eqref{spectrum}, coincide one with another. It happens for  
special values of the integrals of motion characterizing the geodesic trajectories. The 
larger is the number of spacetime dimensions the larger is the number of different degenerate cases. Some of them will be discussed in the next section.

In our setup it is somewhat more natural to construct, 
instead of the vector basis $\{\tens{p}\}$, the parallel-propagated basis of forms $\{\tens{\pi}\}$. In the generic case it consists of  
\begin{equation}
\{\tens{u}^\flat, \tens{z}, \tens{\pi}^{\hat 1}, \tens{\bar \pi}^{\hat 1},\dots,
\tens{\pi}^{\hat n}, \tens{\bar \pi}^{\hat n}\}\,.
\end{equation}
(The element $\tens{z}$ is present only in even dimensions.)
If $\{ \bs{\varsigma}^{\hat \mu},\bs{\bar \varsigma}^{\hat
\mu}\}$ are comoving basis forms spanning $V_\mu^*$, then (cf. Eq. \eqref{pp})
\begin{equation}\label{PTvectors}
\begin{split}
\tens{\pi}^{\hat \mu}=\tens{\varsigma}^{\hat \mu}
\cos\beta_\mu-\tens{\bar \varsigma}^{\hat \mu}\sin\beta_\mu\,,\\ 
\tens{\bar \pi}^{\hat \mu}=\tens{\varsigma}^{\hat \mu}
\sin\beta_\mu+\tens{\bar \varsigma}^{\hat \mu}\cos\beta_\mu\,,
\end{split}
\end{equation}
where
\begin{equation}\label{beta_dot}
{\dot \beta_\mu}=(\tens{\bar \varsigma}^{\hat \mu},
\tens{\dot \varsigma}^{\hat \mu})=
-(\tens{\varsigma}^{\hat \mu}, \tens{\dot{\bar \varsigma}}^{\,\hat \mu} )\,,
\end{equation}
with the initial condition $\beta_\mu(\tau=0)=0\,$.

The rotation angles $\dot \beta_\mu$ as given by \eqref{beta_dot} are functions of 
$r$ and $x_\mu$. In the case when $\dot \beta_\mu$ can be 
brought  into the form
\begin{equation}\label{f}
\dot\beta_\mu=\frac{f^{(\mu)}_n(r)}{U_n}+
\sum_{\nu=1}^{n-1} \frac{f^{(\mu)}_\nu(x_\nu)}{U_\nu}\,,
\end{equation}
the problem \eqref{beta_dot} is {\em separable} and the particular solution 
is given by (see Appendix B)
\begin{equation}\label{separ}
\beta_\mu=\!\int\!\!\frac{\sigma_n f_n^{(\mu)}dr}{\sqrt{W_n^2-X_n V_n}}
-\sum_{\nu=1}^{n-1}\int\!\!\frac{\sigma_\nu f_\nu^{(\mu)}dx_\nu}{\sqrt{X_\nu V_\nu-W_\nu^2}}\,.
\end{equation}

\section{Examples}
We shall now illustrate the above described formalism  by considering  
$D=3, 4, 5$ Kerr-NUT-(A)dS spacetimes.
We take the normalization of the velocity $\kappa_0=-1$
and normalize  other vectors of the
parallel-transported frame to $+1$.
In the derivation of the equations for 
$\dot \beta_{\mu}$ we used the Maple program.
  
\subsection{3D spacetime: BTZ black holes}
\subsubsection{Generic case}

As the first example we consider the case when $D=3$, that is when the metric
\eqref{metrics} describes a BTZ black hole \cite{BTZ}.  We 
first discuss the generic case, $q_0=0$, and then briefly
mention what happens for the degenerate geodesics with $q_0=2$. Since in three dimensions $n=1$ we
drop everywhere index $\mu$.

So, we have the metric
\begin{equation}
\tens{g}=-\tens{\bar \omega}\tens{\bar \omega}+\tens{\omega}\tens{\omega}
+\tens{\omega}^{\hat \epsilon}\tens{\omega}^{\hat \epsilon}\,,
\end{equation}
where  
\ba\label{triad}
\tens{\bar \omega}\!&=&\!\sqrt{X} \tens{d}\psi_0\,,\ \tens{\omega}
\!=\!\frac{\tens{d}r}{\sqrt{X}}\,,\ 
\tens{\omega}^{\hat \epsilon}\!=
\!\frac{\sqrt{c}}{r}(\tens{d}\psi_0\!-r^2 \tens{d}\psi_1 )\,,\nonumber\\
X\!&=&\! c_1 r^2-2m+\frac{c}{r^2}\,.
\ea
The parameter $c_1$ is proportional to the cosmological constant and parameters 
$m$ and $c>0$ are related to mass and rotation parameter.
    
The principal CKY tensor and the Killing tensor are:
\begin{equation}\label{KY}
\tens{h}=-r\tens{\omega}\wedge\tens{\bar \omega}\,,
\quad \tens{K}=-r^2\tens{\omega}^{\hat \epsilon}\tens{\omega}^{\hat \epsilon}\,.
\end{equation}
The velocity
\be
\tens{u}^\flat={\bar u} \tens{\bar \omega}+u \tens{\omega}+
u_{\hat \epsilon} \tens{\omega}^{\hat \epsilon}
\ee
has the components
\be\label{u}
\bar u =\frac{W}{\sqrt{X}}\,,\  
u=\sigma\sqrt{\frac{W^2}{X}-V}\,,\  
u_{\hat \epsilon}=\frac{\Psi_1}{\sqrt{c}r}\,,
\ee
where
\begin{equation}
W=-\Psi_0-\frac{\Psi_1}{r^2}\,,\quad
V=1+\frac{\Psi_1^2}{cr^2}\,.
\end{equation}
In the velocity adapted frame $\{\tens{\bar o},\tens{o}, \tens{o}^{\hat \epsilon}\}$ given by \eqref{newframe} we have
\ba
\tens{u}^\flat\!&=&\!{\bar v}\tens{\bar o}+u_{\hat \epsilon}\tens{o}^{\hat \epsilon}\,,\quad \bar v =-\sqrt{V}\,,\\
\tens{F}\!&=&\!ru_{\hat 0}\tens{o}\wedge
(u_{\hat 0}\tens{\bar o}+\bar v\tens{o}^{\hat \epsilon})\,.
\ea
The spectrum \eqref{spectrum} of $\tens{F}$ is 
\begin{equation}
S(\tens F)=\{0, \lambda ,\lambda\}\,,\quad
\lambda=\frac{|\Psi_1|}{\sqrt{c}}\,.
\end{equation}
The zero eigenvalue corresponds to the space $U^*$ spanned by $\tens{u}^\flat$.
In the non-degenerate case, that is when $\Psi_1\neq 0$, 
the eigenspace $V_0^*$ is trivial.
The orthonormal forms spanning $V_{\lambda}^*$ are:
\begin{equation}
\tens{\varsigma}=\tens{o}\,,\quad 
\tens{\bar \varsigma}=u_{\hat \epsilon} \tens{\bar o}+\bar v \tens{o}^{\hat \epsilon}\,.
\end{equation}
Using \eqref{beta_dot} one finds 
\begin{equation}
\dot \beta=\frac{M}{r^2+\lambda^2}\,,\quad M=\frac{c-\Psi_0\Psi_1}{\sqrt{c}}\,.
\end{equation}
The parallel-transported forms $(\tens{\pi}, \tens{\bar \pi})$ are given by 
\eqref{PTvectors}, where
\begin{equation}\label{beta3d}
\beta=\!\! \int\!\! \frac{\sigma M dr}{(r^2+\lambda^2)\sqrt{W^2-XV}}\,. 
\end{equation}

\subsubsection{Degenerate case}
Let us now consider special geodesic trajectories with 
$\Psi_1=0$ for which $q_0=2$. For such trajectories one has
\be\label{u}
\bar u =-\frac{\Psi_0}{\sqrt{X}}\,,\quad  
u=\sigma\sqrt{\frac{\Psi_0^2}{X}-1}\,,\quad  
u_{\hat \epsilon}=0\,.
\ee
In the adapted basis the velocity is $\tens{u}^\flat=-\tens{\bar o}\,.$ 
Operators $\tens{F}$ and $\tens{F}^2$ become trivial.  
The space $V_0^*$ is spanned by $\{\fb{{1}}{\varsigma}{0}, \fb{{2}}{\varsigma}{0}\}$, where
\begin{equation}
\fb{{1}}{\varsigma}{0}=\tens{o},\quad 
\fb{{2}}{\varsigma}{0}=-\tens{o}^{\hat \epsilon}\,.
\end{equation}
Similar to \eqref{PTvectors} and \eqref{beta_dot} parallel-transported 
forms can be written as follows
\ba\label{PTvectorsDeg}
\tens{\pi}\!&=&\!\fb{{1}}{\varsigma}{0}\cos\beta
-\fb{{2}}{\varsigma}{0}\sin\beta\,,\nonumber\\ 
\tens{\bar \pi}\!&=&\!\fb{{1}}{\varsigma}{0}\sin\beta
+\fb{{2}}{\varsigma}{0}\cos\beta\,,\\
{\dot \beta}\!&=&\!(\fb{{2}}{\varsigma}{0}, \fb{{1}}{\dot \varsigma}{0})=
-(\fb{{1}}{\varsigma}{0}, \fb{{2}}{\dot \varsigma}{0})\,,\nonumber
\ea
with the initial condition $\beta(\tau=0)=0\,$.
Using these equations we find $\dot \beta=\sqrt{c}/r^2$ and hence
\begin{equation}
\beta=\!\! \int\!\! \frac{\sigma \sqrt{c} dr}{r^2\sqrt{\Psi_0^2-X}}\,.
\end{equation}
Notice that this relation can be obtained from \eqref{beta3d} by taking the 
limit $\Psi_1\to 0$. 

To conclude, the parallel-propagated orthonormal frame around a BTZ black hole
is $\{\tens{u}^\flat, \tens{\pi}, \tens{\bar \pi}\}$. This frame remains parallel-propagated 
also off-shell, when $X$ given by \eqref{triad} becomes an arbitrary function 
of $r$.

\subsection{4D spacetime: Carter's family of solutions}
Let us now consider the case of $D=4$. We have
\begin{equation}\label{g4d}
\tens{g}=-\tens{\bar \omega}^{\hat 2}\tens{\bar \omega}^{\hat 2}+\tens{\omega}^{\hat 2}\tens{\omega}^{\hat 2}+
\tens{\bar \omega}^{\hat 1}\tens{\bar \omega}^{\hat 1}+\tens{\omega}^{\hat 1}\tens{\omega}^{\hat 1}\,,
\end{equation}
where  
\begin{equation}\label{tetrad}
\begin{split}
\tens{\bar \omega}^{\hat 2}=&\,\sqrt{\frac{X_2}{U_2}}(\tens{d}\psi_0+x_1^2\tens{d}\psi_1 )\,,\ 
\tens{\omega}^{\hat 2}=\,\sqrt{\frac{U_2}{X_2}}\,\tens{d}r\,,\\ 
\tens{\bar \omega}^{\hat 1}=&\,\sqrt{\frac{X_1}{U_1}}(\tens{d}\psi_0-r^2\tens{d}\psi_1 )\,,\ 
\tens{\omega}^{\hat 1}=\sqrt{\frac{U_1}{X_1}}\, \tens{d}x_1\,. 
\end{split}
\end{equation}
Here, $U_2=-U_1=x_1^2+r^2$,
and we shall not be specifying functions $X_1(x_1), X_2(r)$ at this point. 

The principal CKY tensor and the Killing tensor are:
\ba
\tens{h}\!\!&=&\!\!x_1\tens{\omega}^{\hat 1}\wedge\tens{\bar \omega}^{\hat 1}-r\tens{\omega}^{\hat 2}\wedge\tens{\bar \omega}^{\hat 2}\,,\label{KY4}\\
\tens{K}\!\!&=&\!\!x_1^2(\tens{\omega}^{\hat 2}\tens{\omega}^{\hat 2}\!\!-\!\tens{\bar \omega}^{\hat 2}\tens{\bar \omega}^{\hat 2})
\!-\!r^2(\tens{\bar \omega}^{\hat 1}\tens{\bar \omega}^{\hat 1}\!\!+\!\tens{\omega}^{\hat 1}\tens{\omega}^{\hat 1}).\label{KT4}
\ea
The components of the velocity are 
\be\label{u4}
\begin{split}
{\bar u}_{\hat 2}=&\,\frac{W_2}{\sqrt{X_2U_2}}\,,\  
u_{\hat 2}=\frac{\sigma_2}{\sqrt{X_2U_2}}\sqrt{W_2^2-X_2V_2}\,,\\  
{\bar u}_{\hat 1}=&\,\frac{W_1}{\sqrt{X_1U_1}}\,,\  
u_{\hat 1}=\frac{\sigma_1}{\sqrt{X_1U_1}}\sqrt{X_1V_1-W_1^2}\,, 
\end{split}
\ee
where
\begin{equation}
\begin{split}
W_2=&\,-r^2\Psi_0-\Psi_1\,,\quad
V_2=r^2-\kappa_1\,,\\
W_1=&\,-x_1^2\Psi_0+\Psi_1\,,\quad
V_1=x_1^2+\kappa_1\,.
\end{split}
\end{equation}
The constants of geodesic motion $\Psi_0$ and $\Psi_1$ are associated with isometries and $\kappa_1<0$ corresponds to the Killing tensor \eqref{KT4}.
In the velocity adapted frame $\{\tens{\bar o}^{\hat 2},\tens{o}^{\hat 2},\tens{\bar o}^{\hat 1},\tens{o}^{\hat 1}\}$ given by \eqref{newframe} we have 
\ba
\tens{u}^\flat\!\!\!&=&\!\!{\bar v}_{\hat 2}\tens{\bar o}^{\hat 2}\!\!+{\bar v}_{\hat 1}\tens{\bar o}^{\hat 1},\,{\bar v}_{\hat 2}\!=\!-\sqrt{\frac{V_2}{U_2}}\,,\,{\bar v}_{\hat 1}\!=\!\sqrt{\frac{V_1}{U_1}}\,,\\
\tens{F}\!\!\!&=&\!\!(r{\bar v}_{\hat 1}\tens{o}^{\hat 2}+x_1{\bar v}_{\hat 2}\tens{o}^{\hat 1})\wedge ({\bar v}_{\hat 1}\tens{\bar o}^{\hat 2}+{\bar v}_{\hat 2}\tens{\bar o}^{\hat 1})\,.
\ea
The spectrum \eqref{spectrum} of $\tens{F}$ is
\begin{equation}
S(\tens F)=\{0, 0, \lambda ,\lambda\}\,,\quad
\lambda=\sqrt{-\kappa_1}\,.
\end{equation}
The first zero eigenvalue corresponds to $U^*$, while the second one corresponds 
to the eigenspace $V_0^*$, spanned by $1$-form $\tens{z}$ \eqref{z}. 
When normalized $\tens{z}$ reads:
\begin{equation}\label{z4D}
\tens{z}=\lambda^{-1}(-x_1{\bar v}_{\hat 2}\tens{o}^{\hat 2}+r{\bar v}_{\hat 1}\tens{o}^{\hat 1})\,.
\end{equation}
The orthonormal forms spanning $V_{\lambda}^*$ are:
\begin{equation}
\tens{\varsigma}={\bar v}_{\hat 1} \tens{\bar o}^{\hat 2}+{\bar v}_{\hat 2}\tens{\bar o}^{\hat 1}\,,\ 
\tens{\bar \varsigma}=\lambda^{-1}(r{\bar v}_{\hat 1}\tens{o}^{\hat 2}+x_1{\bar v}_{\hat 2}\tens{o}^{\hat 1})\,.
\end{equation}
Using \eqref{beta_dot} one finds
\ba\label{f1f24D}
\dot \beta\!&=&\!\frac{M}{(x_1^2-\lambda^2)(r^2+\lambda^2)}=
\frac{f_1}{U_1}+\frac{f_2}{U_2}\,,\nonumber\\
f_1\!&=&\!-\frac{M}{x_1^2-\lambda^2}\,,\quad 
f_2=\frac{M}{r^2+\lambda^2}\,,
\ea
where $M=\lambda\bigl(\Psi_1-\lambda^2\Psi_0\bigr)$.
Therefore, $\beta$ allows a separation of variables and can be written as
\begin{equation}\label{beta4D}
\beta=\!\!\int\!\!\frac{\sigma_2 f_2 dr}{\sqrt{W_2^2-X_2V_2}}-
\!\!\int\!\!\frac{\sigma_1 f_1 dx_1}{\sqrt{X_1V_1-W_1^2}}\ ,
\end{equation}
where functions $f_1$, $f_2$ are defined in \eqref{f1f24D}.
The parallel-transported forms $\{\tens{\pi}, \tens{\bar \pi}\}$
are given by \eqref{PTvectors}. 

To summarize, the parallel-propagated orthonormal frame in the spacetime
\eqref{g4d}-\eqref{tetrad} is $\{\tens{u}^\flat, \tens{z}, \tens{\pi}, \tens{\bar \pi}\}$. 
This parallel-propagated basis is constructed for arbitrary functions $X_1(x_1), X_2(r)$, and in particular 
for the Carter's class of solutions \cite{Ca1, Ca2}---describing among others a rotating charged 
black hole in the cosmological background. So we have re-derived the results 
obtained earlier in \cite{Marck, KM}.

\subsection{5D Kerr-NUT-(A)dS spacetime}
\subsubsection{Generic case}
As the last example we consider the 5D off-shell Kerr-NUT-(A)dS spacetime.
Similar to the 3D case we shall first obtain the parallel-propagated frame 
for generic geodesics and then briefly discuss what happens for the special trajectories characterized by $q_0=2$, or $q_1=2$.
The metric reads
\begin{equation}\label{g5d}
\tens{g}=-\tens{\bar \omega}^{\hat 2}\tens{\bar \omega}^{\hat 2}+\tens{\omega}^{\hat 2}\tens{\omega}^{\hat 2}+
\tens{\bar \omega}^{\hat 1}\tens{\bar \omega}^{\hat 1}+\tens{\omega}^{\hat 1}
\tens{\omega}^{\hat 1}
+\tens{\omega}^{\hat \epsilon}\tens{\omega}^{\hat \epsilon}\,,
\end{equation}
where  
\begin{equation}\label{quintad}
\begin{split}
\tens{\bar \omega}^{\hat 2}=&\,\sqrt{\frac{X_2}{U_2}}(\tens{d}\psi_0+x_1^2\tens{d}\psi_1 )\,,\  
\tens{\omega}^{\hat 2}=\,\sqrt{\frac{U_2}{X_2}}\,\tens{d}r\,,\\ 
\tens{\bar \omega}^{\hat 1}=&\,\sqrt{\frac{X_1}{U_1}}(\tens{d}\psi_0-r^2\tens{d}\psi_1 )\,,\  
\tens{\omega}^{\hat 1}=\sqrt{\frac{U_1}{X_1}}\, \tens{d}x_1\,,\\
\tens{\omega}^{\hat \epsilon}=&\,\frac{\sqrt{c}}{rx_1}\,\left[\tens{d}\psi_0+
(x_1^2-r^2)\tens{d}\psi_1-x_1^2r^2\tens{d}\psi_2\right]\,,
\end{split}
\end{equation}
and $U_2=-U_1=x_1^2+r^2$.
The principal CKY tensor and the Killing tensor for this metric are:
\ba\label{KY5}
\tens{h}\!&=&\!x_1\tens{\omega}^{\hat 1}\wedge\tens{\bar \omega}^{\hat 1}-r\tens{\omega}^{\hat 2}\wedge\tens{\bar \omega}^{\hat 2}\,,\\
\tens{K}\!&=&\!x_1^2(-\tens{\bar \omega}^{\hat 2}\tens{\bar \omega}^{\hat 2}+\tens{\omega}^{\hat 2}\tens{\omega}^{\hat 2}
+\tens{\omega}^{\hat \epsilon}\tens{\omega}^{\hat \epsilon})\nonumber\\
&&\!-r^2(\tens{\bar \omega}^{\hat 1}\tens{\bar \omega}^{\hat 1}+\tens{\omega}^{\hat 1}\tens{\omega}^{\hat 1}
+\tens{\omega}^{\hat \epsilon}\tens{\omega}^{\hat \epsilon})\,.
\ea
The components of the velocity are 
\ba\label{u5}
{\bar u}_{\hat 2}\!&=&\!\frac{W_2}{\sqrt{X_2U_2}}\,,\  
u_{\hat 2}=\frac{\sigma_2}{\sqrt{X_2U_2}}\sqrt{W_2^2-X_2V_2}\,,\nonumber\\  
{\bar u}_{\hat 1}\!&=&\!\frac{W_1}{\sqrt{X_1U_1}}\,,\  
u_{\hat 1}=\frac{\sigma_1}{\sqrt{X_1U_1}}\sqrt{X_1V_1-W_1^2}\,,\nonumber\\
u_{\hat \epsilon}\!&=&\!\frac{\Psi_2}{\sqrt{c}x_1r}\,, 
\ea
where
\begin{equation}\label{W5}
\begin{split}
W_1=&\,-x_1^2\Psi_0+\Psi_1-\frac{\Psi_2}{x_1^2}\,,\ 
V_1=x_1^2+\kappa_1+\frac{\Psi_2^2}{cx_1^2}\,,\\
W_2=&\,-r^2\Psi_0-\Psi_1-\frac{\Psi_2}{r^2}\,,\ 
V_2=r^2-\kappa_1+\frac{\Psi_2^2}{cr^2}\,.
\end{split}
\end{equation}
In the velocity adapted frame $\{\tens{o}\}$ given by \eqref{newframe}
we have 
\ba\label{u5c}
\tens{u}^\flat\!&=&\!{\bar v}_{\hat 2}\tens{\bar o}^{\hat 2}\!+{\bar v}_{\hat 1}\tens{\bar o}^{\hat 1}\!+u_{\hat \epsilon}\tens{o}^{\hat \epsilon}\,,\nonumber\\ 
{\bar v}_{\hat 2}\!&=&\!-\sqrt{\frac{V_2}{U_2}}\,,\quad 
{\bar v}_{\hat 1}=\sqrt{\frac{V_1}{U_1}}\,.
\ea
The form $\tens{F}$ is
\begin{equation}\label{F5D}
\begin{split}
\tens{F}=&\,(r{\bar v}_{\hat 1}\tens{o}^{\hat 2}+x_1{\bar v}_{\hat 2}\tens{o}^{\hat 1})\wedge
({\bar v}_{\hat 1}\tens{\bar o}^{\hat 2}+{\bar v}_{\hat 2}\tens{\bar o}^{\hat 1})\\
+&\,ru_{\hat \epsilon}\tens{o}^{\hat 2}\wedge(\bar{v}_{\hat 2}\tens{o}^{\hat \epsilon}+u_{\hat \epsilon}\tens{\bar o}^{\hat 2})\\
+&\,x_1u_{\hat \epsilon}\tens{o}^{\hat 1}\wedge({\bar v}_{\hat 1}\tens{o}^{\hat \epsilon}-u_{\hat \epsilon}\tens{\bar o}^{\hat 1})\,.
\end{split}
\end{equation}
The 2-form $\tens{z}$ \eqref{z} reads
\begin{equation}\label{z5D}
\tens{z}\!=\!\tens{o}^{\hat \epsilon}\!\wedge (r{\bar v}_{\hat 1}\tens{o}^{\hat 1}\!-
x_1{\bar v}_{\hat 2}\tens{o}^{\hat 2})\!+u_{\hat \epsilon}(x_1\tens{o}^{\hat 2}\!\wedge
\tens{\bar o}^{\hat 2}\!+r\tens{o}^{\hat 1}\!\wedge\tens{\bar o}^{\hat 1})\,.
\end{equation}
The spectrum \eqref{spectrum} of $\tens{F}$ is
\begin{equation}
S(\bs F)=\{0, \lambda_1 ,\lambda_1, \lambda_2, \lambda_2\}\,,
\end{equation}
where 
\begin{equation}
\lambda_1^2\!=\!-\frac{\!\kappa_1\!+\!D}{2}\,,\,  
\lambda_2^2\!=\!-\frac{\!\kappa_1\!-\!D}{2}\,,\, 
D\!=\!\sqrt{\kappa_1^2\!-\!4\frac{\Psi_2^2}{c}}\,.
\end{equation}
The zero eigenvalue corresponds to $U^*$.
The eigenspace $V_{1}^*$ is spanned by 
\ba
\tens{\varsigma}^{\hat 1}\!&=&\!N_1\Bigl(-\frac{x_1 F_r(\lambda_2)}{\sqrt{U_2}}\,\tens{\bar o}^{\hat 2}+\frac{rF_{x_1}(\lambda_2)}{\sqrt{U_2}}\,\tens{\bar o}^{\hat 1}+\tens{o}^{\hat \epsilon}\Bigr)\,,\nonumber\\
\tens{\bar \varsigma}^{\hat 1}\!&=&\!{\bar N}_{1}\Bigl(F_r(\lambda_2)F_{x_1}(\lambda_1)\tens{o}^{\hat 2}+\tens{o}^{\hat 1}\Bigr)\,.
\ea
Here $N_1$ and ${\bar N}_{1}$ stand for normalizations,
\ba
N_1\!&=&\!\sqrt{U_2}/\sqrt{U_2+r^2F_{x_1}(\lambda_2)^2-x_1^2F_r(\lambda_2)^2}\,,\nonumber\\
{\bar N}_{1}\!&=&\!1/\sqrt{1+F_{x_1}(\lambda_1)^2F_r(\lambda_2)^2}\,,
\ea
and we have introduced
\begin{equation}
F_r(\lambda)=\frac{x_1u_{\hat \epsilon}\sqrt{V_2}}{x_1^2u_{\hat \epsilon}^2+\lambda^2}\,,\ 
F_{x_1}(\lambda)=\frac{ru_{\hat \epsilon}\sqrt{-V_1}}{r^2u_{\hat \epsilon}^2-\lambda^2}\,, 
\end{equation}
which are functions of $r$, $x_1$, respectively.
Using \eqref{beta_dot} we find
\begin{equation}
\begin{split}
\dot \beta_{1}=&\,\frac{M^{(1)}}{(x_1^2-\lambda_1^2)(r^2+\lambda_1^2)}=\frac{f_1^{(1)}}{U_1}+\frac{f_2^{(1)}}{U_2}\,,\\
f_1^{(1)}=&\,-\frac{M^{(1)}}{x_1^2-\lambda_1^2}\,,\quad 
f_2^{(1)}=\frac{M^{(1)}}{r^2+\lambda_1^2}\,,
\\
M^{(1)}\!=&\,-\frac{\Psi_2\Psi_1-\lambda_1^2\Psi_2\Psi_0-c\lambda_2^2}{\lambda_2\sqrt{c}}\,.
\end{split}
\end{equation}
This means that $\beta_1$ can be separated:
\begin{equation}\label{beta1}
\beta_1=\!\int\!\!\frac{\sigma_2 f_2^{(1)} dr}{\sqrt{W_2^2-X_2 V_2}}
-\!\int\!\!\frac{\sigma_{1} f_1^{(1)} dx_1}{\sqrt{X_1 V_1-W_1^2}}\,.
\end{equation} 
The parallel-propagated forms $\{\tens{\pi}^{\hat 1}, \tens{\bar \pi}^{\hat 1}\}$ are given by rotation \eqref{PTvectors}.
Similarly one finds that
\ba
\tens{\varsigma}^{\hat 2}\!&=&\!N_2\Bigl(-\frac{x_1 F_r(\lambda_1)}{\sqrt{U_2}}\,\tens{\bar o}^{\hat 2}+\frac{rF_{x_1}(\lambda_1)}{\sqrt{U_2}}\,\tens{\bar o}^{\hat 1}+\tens{o}^{\hat \epsilon}\Bigr)\,,\nonumber\\
N_2\!&=&\!\sqrt{U_2}/\sqrt{U_2+r^2F_{x_1}(\lambda_1)^2-x_1^2F_r(\lambda_1)^2}\,,\nonumber\\
\tens{\bar \varsigma}^{\hat 2}\!&=&\!{\bar N}_{2}\Bigl(F_r(\lambda_1)F_{x_1}(\lambda_2)\tens{o}^{\hat 2}+\tens{o}^{\hat 1}\Bigr)\,,\nonumber\\
{\bar N}_{2}\!&=&\!1/\sqrt{1+F_{x_1}(\lambda_2)^2F_r(\lambda_1)^2}\,,
\ea
span the eigenspace $V_{2}^*$.
Using \eqref{beta_dot} we have
\begin{equation}
\begin{split}
\dot \beta_{2}=&\,\frac{M^{(2)}}{(x_1^2-\lambda_2^2)(r^2+\lambda_2^2)}=\frac{f_1^{(2)}}{U_1}+\frac{f_2^{(2)}}{U_2}\,,\\
f_1^{(2)}=&\,-\frac{M^{(2)}}{x_1^2-\lambda_2^2}\,,\quad 
f_2^{(2)}=\frac{M^{(2)}}{r^2+\lambda_2^2}\,,
\\
M^{(2)}=&\,\frac{\Psi_2\Psi_1-\lambda_2^2\Psi_2\Psi_0-c\lambda_1^2}{\lambda_1\sqrt{c}}\,.
\end{split}
\end{equation}
The parallel-propagated forms $\{\tens{\pi}^{\hat 2}, \tens{\bar \pi}^{\hat 2}\}$ 
are given by rotation \eqref{PTvectors}, where
\begin{equation}\label{beta2}
\beta_2=\!\int\!\!\frac{\sigma_2 f_2^{(2)} dr}{\sqrt{W_2^2-X_2 V_2}}
-\!\int\!\!\frac{\sigma_{1} f_1^{(2)} dx_1}{\sqrt{X_1 V_1-W_1^2}}\,.
\end{equation}

\subsubsection{Degenerate case}
In $D=5$ two different degenerate cases are possible.
One can have either a 2-dimensional $V_0$ ($q_0=2$) which happens for 
the special geodesics characterized by 
$\Psi_2=0$, or a 4-dimensional $V_\lambda$ ($q_1=2$)
which happens when $\kappa_1^2=4\Psi_2^2/c\,.$
The latter case is more complicated and the general formulas \eqref{tran}-\eqref{NN}
have to be used. We shall not do this here and rather concentrate 
on the first degeneracy which has an interesting consequence.

So we consider the special geodesics characterized by $\Psi_2=0$.
It can be checked by direct calculations that 
in this case the results can be obtained by taking the 
limit $\Psi_2\to 0$ of previous formulas. 
In particular one has $u_{\hat \epsilon}=0$,
\begin{equation}\label{W5d}
\begin{split}
W_1=&\,-x_1^2\Psi_0+\Psi_1\,,\quad 
V_1=x_1^2+\kappa_1\,,\\
W_2=&\,-r^2\Psi_0-\Psi_1\,,\quad 
V_2=r^2-\kappa_1\,.
\end{split}
\end{equation}
The velocity $\tens{u}$ becomes effectively 4-dimensional:
\begin{equation}
\tens{u}^\flat={\bar v}_{\hat 2}\tens{\bar o}^{\hat 2}+{\bar v}_{\hat 1}\tens{\bar o}^{\hat 1}\,.
\end{equation}
The form $\tens{F}$, \eqref{F5D}, becomes degenerate and takes the form
\begin{equation}
\begin{split}
\tens{F}=&\,(r{\bar v}_{\hat 1}\tens{o}^{\hat 2}+x_1{\bar v}_{\hat 2}\tens{o}^{\hat 1})\wedge
({\bar v}_{\hat 1}\tens{\bar o}^{\hat 2}+{\bar v}_{\hat 2}\tens{\bar o}^{\hat 1})\,.
\end{split}
\end{equation}
The 2-form $\tens{z}$ \eqref{z5D} reduces to 
\begin{equation}
\tens{z}\!=\!\tens{o}^{\hat \epsilon}\!\wedge (r{\bar v}_{\hat 1}\tens{o}^{\hat 1}-
x_1{\bar v}_{\hat 2}\tens{o}^{\hat 2})\,.
\end{equation}
The spectrum is 
\begin{equation}
S(\bs F)=\{ 0, 0, 0, \lambda, \lambda\}\,,\quad 
\lambda=\sqrt{-\kappa_1}\,.
\end{equation}
The eigenspace $V_0^*$ is spread by 
$\{\fb{{1}}{\varsigma}{0}, \fb{{2}}{\varsigma}{0}\}$, where
\begin{equation}
\fb{{1}}{\varsigma}{0}=\tens{o}^{\hat \epsilon},\quad  
\fb{{2}}{\varsigma}{0}=\lambda^{-1}(r{\bar v}_{\hat 1}\tens{o}^{\hat 1}-
x_1{\bar v}_{\hat 2}\tens{o}^{\hat 2})\,.
\end{equation}
(Notice that $\fb{{2}}{\varsigma}{0}$ is identical to the normalized 4-dimensional 1-form $\tens{z}$ given by \eqref{z4D}.) 
The angle of rotation in the $\{\fb{{1}}{\varsigma}{0}, \fb{{2}}{\varsigma}{0}\}$ 2-plane obeys
the equation
\ba\label{beta1Deg}
\dot \beta_1\!\!&=&\!\frac{M^{(1)}}{x_1^2r^2}
=\frac{f_1^{(1)}}{U_1}+\frac{f_2^{(1)}}{U_2}\,,\ 
M^{(1)}=\lambda\sqrt{c}\,,\nonumber\\
f_1^{(1)}\!\!&=&\!-\frac{M^{(1)}}{x_1^2}\,,\quad  
f_2^{(1)}=\frac{M^{(1)}}{r^2}\,.
\ea
Thus $\beta_1$ is given by \eqref{beta1}
with functions $f_1^{(1)}, f_2^{(1)}$ defined in \eqref{beta1Deg}.
The eigenspace $V_{\lambda}^*$ spread by
\begin{equation}
\tens{\varsigma}={\bar v}_{\hat 1}\tens{\bar o}^{\hat 2}+{\bar v}_{\hat 2}\tens{\bar o}^{\hat 1}\,,\ 
\tens{\bar \varsigma}=\lambda^{-1}(r{\bar v}_{\hat 1}\tens{o}^{\hat 2}+x_1{\bar v}_{\hat 2}\tens{o}^{\hat 1})\,, 
\end{equation}
is identical to the $V_\lambda^*$ subspace in the 4D case. Thus the rotation angle $\beta_2$ coincides with   
$\beta$ given by \eqref{beta4D}.

\subsubsection{Summary of 5D}

To summarize, we have demonstrated that also in $D=5$ Kerr-NUT-(A)dS
spacetime the rotation angles in 2D eigenspaces can be separated and the
parallel-transported frame $\{\tens{\pi}\}$ explicitly constructed.  This result is
again valid off-shell, that is for arbitrary functions $X_2(r)$,
$X_1(x_1)$.

The special degenerate case characterized by $\Psi_2=0$ 
has the following interesting feature.
The zero-value eigenspace is spanned by the
4-dimensional $\tens{u}^\flat$, by the 4-dimensional  1-form 
$\tens{z}$, and $\tens{o}^{\hat \epsilon}$. The structure of
$V_{\lambda}^*$ is identical to the 4D case and the equation of
parallel transport in this plane is identical to the equation in 4D.
Therefore this 5D degenerate problem effectively reduces to
the generic 4D problem plus the rotation in the 2D 
$\{\tens{z}, \tens{o}^{\hat \epsilon}\}$ plane.
   
This indicates that a similar reduction might be valid also in higher dimensions.
Namely, one may expect that 
the degenerate odd dimensional problem, $\Psi_n=0$,  effectively reduces to the
problem in a spacetime of one dimension lower plus the rotation  in the 2D
$(\tens{z}, \tens{o}^{\hat \epsilon})$ plane.  
If this is so, one can use this odd dimensional degenerate case to generate 
the solution for the generic (one dimension lower) even dimensional problem.

\section{Conclusions} 
 
In this paper we have described the construction of a
parallel-transported frame in a spacetime admitting the principal
CKY tensor $\bs{h}$. This tensor determines a principal (Darboux)
basis at each point of the spacetime. The geodesic motion of a particle
in such a space can be characterized by the components of its velocity
$\bs{u}$ with respect to this basis. For a moving particle it is 
also natural to introduce a comoving basis, which is just a
Darboux basis of $\bs{F}$, where $\bs{F}$ is a projection of $\bs{h}$
along the velocity $\bs{u}$. Since $\bs{F}$ is parallel-propagated
along $\bs{u}$, its eigenvalues are constant along the
geodesic and its eigenspaces are parallel-propagated. We have
demonstrated that for a generic motion the parallel-propagated
basis can be obtained from the comoving basis by simple
two-dimensional rotations in the 2D eigenspaces of $\bs{F}$. 

To illustrate the general theory we have considered the parallel transport
in the Kerr-NUT-(A)dS spacetimes. Namely, we have newly constructed the
parallel-propagated frames in three and five dimensions and
re-derived the results \cite{Marck,KM} in 4D.  One of the interesting
features of the 4D construction, observed already by Marck, is that
the equation for the rotation angle allows a separation of variables.
Remarkably, we have shown that also in five dimensions equations
for the rotation angles can be solved by a separation of variables.
Moreover, the 4D result  can be understood as a special degenerate case 
of the 5D construction. 
Is this a general property valid in the  Kerr-NUT-(A)dS spacetime
with any number of dimensions? What underlines the
separability of the rotation angles? These are interesting open questions. 

The analysis of the present paper was restricted to the problem of 
parallel transport along timelike geodesics. As we already mentioned, 
the generalization to the case of spatial geodesics is straightforward.
The case of null geodesics is a special one and requires additional
consideration (see \cite{M2, StCo:77,CoSt:77,CoPiSt:80} for the 4D case).
The parallel transport along null geodesics might be
of interest for the study of the polarization propagation of 
massless fields with spin in the geometric optics approximation. 

To conclude the paper we would like to mention that described in the
paper the possibility of solving the parallel transport equations in
the Kerr-NUT-(A)dS spacetime is one more evidence of the miraculous
properties of these metrics connected with their hidden symmetries.

\appendix

\section{Remarks on the degeneracy of eigenvalues of $\bs{F}$}

Consider a {\em strictly non-degenerate} operator $\bs{h}$. Let us
discuss what restrictions this condition imposes on the operator
$\bs{F}$. 
If $\bs{h}$ is strictly non-degenerate, then
\be\n{Del}
\Delta(\lambda)=\mbox{det}(h^a_{\ b}-\lambda \delta^a_b)
=(-\lambda)^{\varepsilon}
\prod_{k=1}^n(\lambda^2+\nu_k^2)\, ,
\ee
where all $\nu_k$ are different.

Let us re-calculate this determinant in terms of $\tens{F}$ and compare the results.
For this calculation we use the  Darboux basis of $\tens{F}$ and 
corresponding matrix form of the objects. In particular, we have 
the expression \eqref{F} for $\tens{F}$, and 
\ba\label{S}
s\!&=&\!(s_{\hat 0}, s_{\hat 1}, \dots s_{\hat p})\,,\ 
s_{\hat 0}=(\vnb{1}{s}{{0}},\dots, \!\vnb{q_0}{s}{{0}})\,,\nonumber\\
s_{\hat \mu}\!&=&\!(\vnb{1}{s}{{\mu}},\vnb{1}{\bar s}{{\mu}},\dots,
\vnb{q_\mu}{s}{{\mu}},\vnb{q_\mu}{\bar s}{{\mu}})\,,
\ea
for the 1-form $\tens{s}$. Using \eq{fh}, \eqref{F}, and \eqref{S},
one can rewrite \eqref{Del} as 
\be
\Delta(\lambda)\!=\mbox{det}(F^a_{\ \,b}\!-\!u^as_b\!+\!s^au_b\!-\!\lambda \delta^a_b)\!=\!\left|  
\begin{array}{cc}
A & B\\
C & E
\end{array}
\right|\, ,
\ee
where $A=-\lambda$, $B=-s$, $C=-s^T$, and $E$ is the $(D-1)$-dimensional  
matrix of the form  
\ba\n{DD}
E&=&\mbox{diag}(-\lambda I_{q_0}, \nbsi{1}{Z}, 
\ldots, \nbsi{p}{Z})
\,,\nonumber\\
\nbsi{\mu}{Z}&=&\left(   
\begin{array}{cc}
-\lambda {I}_\mu & \lambda_\mu I_\mu\\
-\lambda_\mu I_\mu & -\lambda I_\mu
\end{array}
\right)
\, .
\ea
Here $I_{q_0}$  is a unit $q_0\times q_0$ matrix, 
and we use $X^T$ to denote a matrix transposed to $X$. It is easy to check that 
\ba\n{DDD}
E^{-1}\!\!&=&\!\mbox{diag}(-\lambda^{-1}I_{q_0},
 \nbsi{1}{Z}^{-1}, \ldots, \nbsi{p}{Z}^{-1}) \,,\nonumber\\
\nbsi{\mu}{Z}^{-1}\!\!&=&\!Q_\mu^{-1}\ \nbsi{\mu}{Z}^T \,,\\
Q_\mu\!\!&=&\!(\lambda^2+\lambda_\mu^2)\hh \mbox{det}(\nbsi{\mu}{Z})
=Q_\mu^{q_\mu}\,.\nonumber
\ea
One has the following
relation for the determinant of a block matrix (see, e.g.,
\cite{Gant})
\be
\left|   
\begin{array}{cc}
A & B\\
C & E
\end{array}
\right|={\cal A}\,|E|\,,\quad {\cal A}=|A-BE^{-1}C|\, .
\ee
Using \eqref{DD} and \eqref{DDD} one finds
\be\label{cco}
\mbox{det}(E)=(-\lambda)^{q_0}\!\! \prod_{\mu=1}^p Q_\mu^{q_\mu}\,,\quad   
{\cal A}=-\lambda-s {E}^{-1}s^T.
\ee
Combining all these relations one obtains
\ba\label{A8}
\Delta(\lambda)\!\!&=&\!\!(-\lambda)^{q_0-1} \prod_{\mu=1}^p Q_\mu^{q_\mu}\nonumber\\
&\times&\!\left[\lambda^2 \Bigl(1-\sum_{\mu=1}^p
\frac{s_{\hat \mu}^2}{Q_\mu}\Bigr)- s_{\hat 0}^2 \right]\,
,\n{DL}\\
s_{\hat 0}^2\!\!&=&\!\! \sum_{i=1}^{q_0} \vnb{i}{s}{{0}}^2\,,\quad  
s_{\hat \mu}^2=\sum_{i=1}^{q_\mu} (\vnb{i}{s}{{\mu}}^2+\vnb{i}{\bar s}{{\mu}}^2)\,.\nonumber
\ea 

Let us now compare \eqref{Del} and \eqref{DL}.
First of all, let us compare the powers of $(-\lambda)$. 
For $s_{\hat 0}^2\ne 0$ we have match 
for $q_0-1=\varepsilon$, whereas the case $s_{\hat 0}^2=0$ may happen only in 
odd dimensions and one must have $q_0=0$ (cf. \eqref{q0}).
Another result of the comparison is that $q_\mu\le 2$. Really,
if $q_\mu>2$, then at least 2 roots of $\Delta(\lambda)$ in \eqref{A8} coincide. This
contradicts the assumption previously stated, since for a strictly non-degenerate
operator $\bs{h}$ the characteristic polynomial has only single roots
$\lambda^2=-\nu_k^2$. The case when $q_\mu=2$ is degenerate. It is valid
only for a special value of the velocity $\bs{u}$. Really, in this case
one of the eigenvalues, say $\nu_k$, of $\bs{h}$ coincides with one of
the eigenvalues of $\bs{F}$ so that one has
$\mbox{det}(\bs{F}-\nu_k\tens{I})=0$. The latter is an equation
restricting the value of $\bs{u}$.

\section{Separability of rotation angles in Kerr-NUT-(A)dS spacetimes}
The right hand sides of the equations \eqref{beta_dot} for rotation angles $\dot \beta_\mu$  
are in general complicated functions of  $r$ and $x_\mu$.
However, it turns out that, at least in four and five dimensions, one can find a particular 
solution for the rotation angles which allows the additive separation of variables. Let us probe this possibility in more detail. The separability means, that we seek the solution in the form 
\begin{equation}\label{psidot}
\beta_\mu=S^{(\mu)}_n(r)-\sum_{\nu=1}^{n-1} S^{(\mu)}_\nu(x_\nu)\,.
\end{equation}
Using \eqref{one-forms} and \eqref{basisvectors} one finds 
\begin{equation}\label{beta_dot2}
\dot \beta_\mu\!=\!\bigl(S_n^{(\mu)}\bigr)'{\dot r}-\!\!\sum_{\nu=1}^{n-1} \bigl(S^{(\mu)}_\nu\bigr)'{\dot x_\nu}\!=\!
\!\sum_{\nu=1}^{n}\! \frac{\bigl(S^{(\mu)}_\nu\bigr)' \!u_{\hat \nu}
\! \sqrt{X_\nu U_\nu}}{U_\nu}\,.
\end{equation}
Prime denotes the derivative with respect to a single argument.
For each $\nu$ the numerator of the latter expression is function of one variable only.
If $\dot \beta_\mu$ given by \eqref{beta_dot} can be brought  into the form
\begin{equation}\label{f}
\dot\beta_\mu=\frac{f^{(\mu)}_n(r)}{U_n}+\sum_{\nu=1}^{n-1} \frac{f^{(\mu)}_\nu(x_\nu)}{U_\nu}\,,
\end{equation}
the problem is separable. By comparing \eqref{beta_dot2} and \eqref{f}
we arrive at the relation
\begin{equation}\label{g}
\frac{g^{(\mu)}_n(r)}{U_n}+\sum_{\nu=1}^{n-1} \frac{g^{(\mu)}_\nu(x_\nu)}{U_\nu}=0\,,
\end{equation}
where ($\nu=1,\dots,n$)
\be
g^{(\mu)}_\nu=\bigl(S^{(\mu )}_\nu\bigr)' u_{\hat \nu}
\sqrt{X_\nu U_\nu}-f^{(\mu)}_\nu\,.
\ee
The general solution of \eqref{g} is (see, e.g., \cite{Krtous})
\be
g^{(\mu)}_\nu\!=\!\!\sum_{k=1}^{n-1}C_k^{(\mu)}(-x_\nu^2)^{n-1-k}\!,\ \,\nu=1,\dots,n\,.
\ee
However, what we need is a particular solution for which we may choose
all constants $C_k^{(\mu)}=0$. Such a particular solution is 
\begin{equation}\label{separ}
\beta_\mu=\!\int\!\!\frac{\sigma_n f_n^{(\mu)}dr}{\sqrt{W_n^2-X_n V_n}}
-\sum_{\nu=1}^{n-1}\int\!\!\frac{\sigma_\nu f_\nu^{(\mu)}dx_\nu}{\sqrt{X_\nu V_\nu-W_\nu^2}}\,.
\end{equation}

\section*{Acknowledgments}
V.F. thanks the Natural Sciences and Engineering Research Council of Canada and
the Killam Trust for the financial support.
D.K. is grateful to the Golden Bell Jar Graduate
Scholarship in Physics at the University of Alberta.

\end{document}